# Neutron Spectroscopy and Computational Methods in investigation of Na ion Battery Materials: A Perspective

Rastislav Turányi and Sanghamitra Mukhopadhyay*

ISIS Neutron and Muon Source, Science and Technology Facilities Council, UKRI.

*sanghamitra.mukhopadhyay@stfc.ac.uk

Adoption of renewable energy is essential to address the challenge of climate change, but that necessitates energy storage technologies. Lithium-ion batteries, the most ubiquitous solution, are insufficient for large-scale applications, so sodium-ion batteries (SIBs), an alternative, are of great current interest. To design and synthesise a commercially viable SIB with required performance, a fundamental understanding of the materials is imperative. Neutron diffraction and spectroscopy provide a unique insight of atomistic understanding of structure and dynamics in materials, therefore in this perspective we have explored how these techniques have been used in SIB research. As neutrons have high penetrability in materials and neutron-matter interaction probabilities are independent of the atomic number, they can provide unique information about motion of particles in bulk materials in the pico- to nanosecond time scales. This makes neutron scattering techniques important tools in battery research. Sodium has a low neutron cross section, which makes computational simulations essential for analysing neutron scattering data of SIB materials. With the availability of high flux neutron sources, high resolution instruments and high performance computers and simulations tools, neutron spectroscopy has been an emerging technique in the last decade for SIB research. In this perspective, we have shown that neutron diffraction is the most popular, while neutron spectroscopies, are just emerging. Computational simulation methods, both force field based and from first principles, are common but still are mostly used independent of neutron experiments. We have identified that suitable improvements of instrumentation, sample environments and simulations methodology will allow these techniques to be more accessible in future.

## 1. Introduction

Renewable energy sources and storage devices have become both a necessity and an inevitability[1-4] to address the challenge of climate change[1,2] and the imminent depletion of fossil fuels[3]. Electric vehicles are seeing a steep increase in demand and this, together with the proliferation of portable electronic devices, are increasing the requirements for both the amount and quality of rechargeable batteries.[5,6] Lithium-ion batteries (LIBs) are commonly used,[5–7] but while they may be able to fulfil the latter demand, the former has been called into question.[7–9] Estimates vary, but the scarcity and toxicity of lithium and other vital materials such as lithium and cobalt in LIBs have raised concerns over future use. Besides, even the best-case scenarios would be very expensive, let alone the environmental cost of extracting these metals.[7–11]

This is especially challenging when considering large-scale electrochemical storage for grid applications. More renewable energy power plants are indispensable, but the most ubiquitous types, solar and wind, produce intermittently and in an amount much smaller compared to the peak demand.[12,13] Thus, cheap storage with long lifecycle is required. Compounded with the aforementioned issues, LIBs also require expensive copper current collectors, making them altogether too expensive.[15] To address this challenge, non-lithium electrochemical batteries are one of the highly promising solutions.[13,14]. Therefore, in recent years there has been huge research interest in batteries based on alternative ions, especially sodium (Na).[13,15]

The sketch of a typical sodium-ion ($Na^+$) battery (SIB) is shown in Figure 1. SIBs were originally the more promising ones, thanks to the high conductivity of $Na^+$ in β-$Al_2O_3$, which has been used as a solid state electrolyte, but there were few successful commercialisations before LIBs overshadowed them[13]. The interest in SIBs renewed recently due to the abundance of Na, the similarity between lithium (Li) and Na ions, and the increasingly pressing issues with LIBs. Using materials, ideas, and lessons from LIBs, a multitude of SIB materials have been tested, showing that good properties can be obtained with cheap and abundant materials like iron (Fe) and manganese (Mn). Additionally, SIB current collectors are less expensive since they can be made with aluminium (Al). Further, unlike LIBs, SIBs can be stored and transported at discharged state, making the processes cheaper and safer, and since they are similar to LIBs, it should be possible to use existing LIB manufacturing lines for SIBs.[13,15–19]

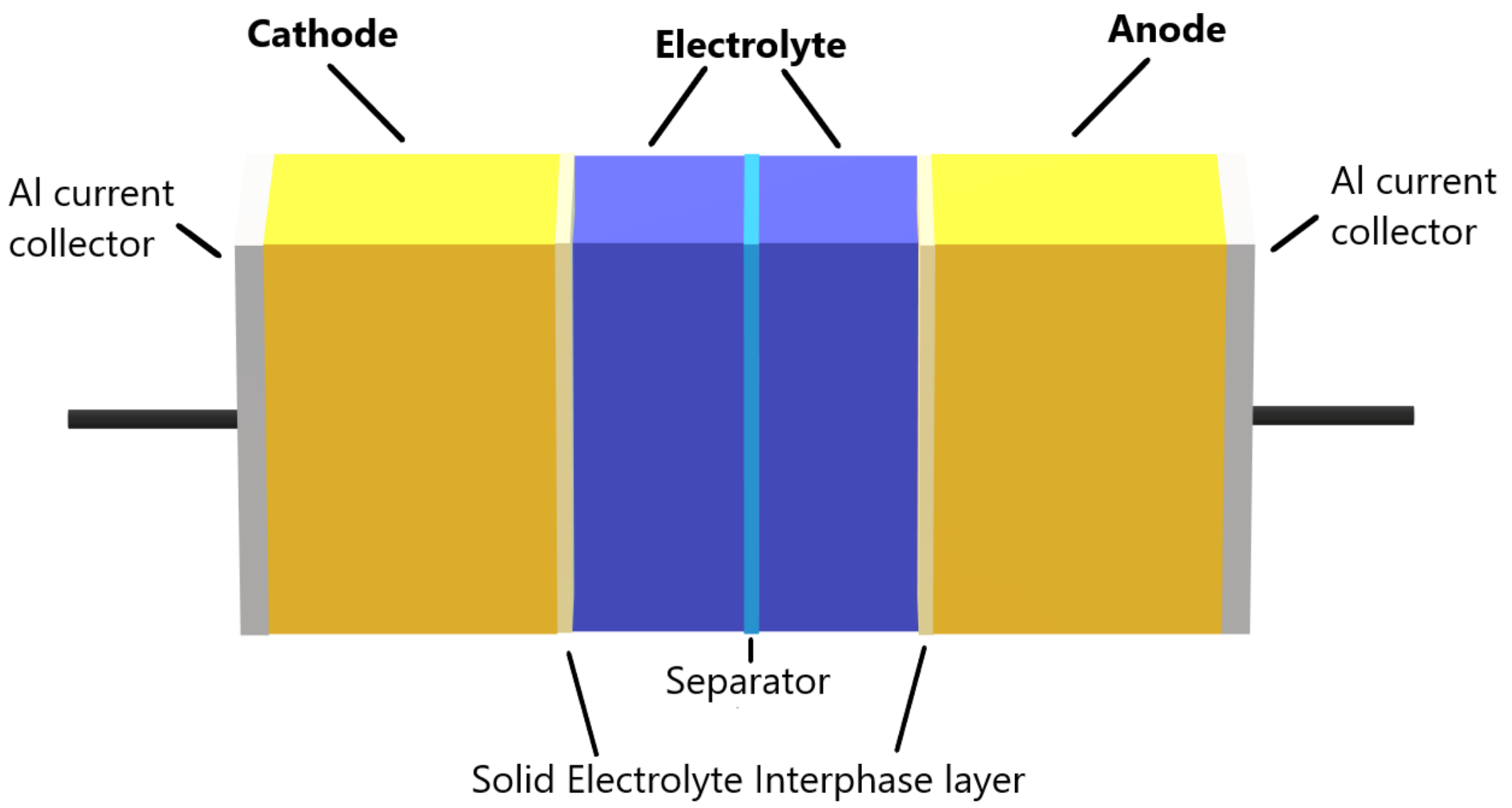

*Figure 1: A diagram of an SIB with its component parts labelled.*

Overall, this shows SIBs as a cheaper, more environmentally friendly alternative to LIBs, but there are challenges too. Sodium equivalents of the most successful LIB electrode materials, such as $LiCoO_2$ cathodes and graphite anodes, provide poor electrochemical performance,[18,20,21] and among other tested materials, none has exhibited an overall performance comparable to LIBs. That said, many properties of SIB materials are comparable to or better than those of LIB materials.[13,15–19] This is because small variations in material structure and synthesis method can have profound effects on performance, allowing for optimisation of different properties.[18,22,23] Therefore, batteries can be prepared for specific applications, such as storage capacities or charging speed. This variety further complicates research and optimisation, necessitating a deep understanding of the materials.[13]

Data from a variety of experimental and computational techniques, performed on a variety of related materials, need to be collated.[16,24] to achieve this understanding. Neutron techniques are an indispensable part of this, thanks to the unique information that they are able to provide, particularly structures and diffusion processes, which will be discussed in the next section.[16,19] This review aims to give an up-to-date account of the uses of experiments complemented with computational simulations in SIB material research.

## 2. Neutron spectroscopy and battery materials

Several characterisation techniques have been used in the last four decades to search for suitable SIB materials. The use of neutron scattering, particularly neutron spectroscopy, however, has been started very recently. This is partially due to the unavailability of suitable neutron sources, but also the low neutron cross section of the Na ion. With the availability of state-of-the-art high energy neutron sources, high resolution, wide dynamic range neutron instruments and high-quality sample environments, it is now possible to use neutron spectroscopy as a characterisation tool for SIB materials. Availability of very high-performance computers and sophisticated computational simulation methods also provide reliable data analysis options to compensate for the issue of low scattering cross section of Na ions.

The reason why these developments are of particular interest lies in the information that neutron techniques are capable of providing. For an overview of the different experimental techniques used in SIB material research, the reader is pointed to the review by Shadike *et al.*,[24] but the specific advantages of neutrons are explored in this section, with the key being complementarity with other techniques.

X-rays interact with electrons and so give stronger signals with heavier atoms and cannot distinguish between atoms with similar atomic numbers. Neutrons, on the other hand, interact with nuclei and so give strong signals with atoms with high neutron scattering cross sections, which includes many light elements. They can also distinguish between neighbouring atoms, and even between isotopes giving an unique opportunity in contrasting different atomic elements and sites. This allows researchers to use techniques such as neutron diffraction (ND) and x-ray diffraction (XRD), as well as neutron pair distribution function (nPDF) and x-ray pair distribution function (xPDF) methods in conjunction to obtain a highly complete structure of a material.[17,24–26] Further, nPDF and nuclear magnetic resonance (NMR) can both be used to investigate local structure.[24]

Inelastic neutron scattering spectroscopy (INS), or neutron vibrational spectroscopy (NVS), is an important characterisation tool complementary to IR and Raman scattering, which can be used to probe the dynamics in these materials. Since INS doesn't have any selection rules, all vibrational frequencies should be observed, in-principles, providing a detailed picture of dynamics. Quasielastic neutron scattering (QENS) spectroscopy can provide unique information about motion of particles in the picosecond to nanosecond time scale. Probing this motion, particularly at high operating temperatures, is utmost important in battery research, because ion migration is an important part in operation of battery.

Neutron spectroscopy has an additional advantage over methods using electromagnetic radiations. Most of techniques using electromagnetic radiations are surface based techniques. On the other hand, neutrons interact only with nuclei and with their high penetrative power, they allow to probe inside the bulk material far from its surface. Apart from that neutron has a physical mass, which helps to find out information about momentum change of moving particles, which is not possible to achieve otherwise using only electromagnetic techniques. The downside of this, however, is that a large amount of sample is required for an experiment, and sample preparation, cell construction, and analysis are much more complex. The result is that neutron experiments are complex to perform but with great potential benefits. This is especially pronounced in *in situ* and *operando* experiments where cells and operating environment during charging-discharging cycle are highly important.[16,17,24–26]

Computational methods provide an additional powerful tool complemented the neutron spectroscopy. Both lattice dynamics and molecular dynamics are useful in analysis of INS and quasi-elastic neutron spectroscopy (QENS), respectively. Apart from these data analysis, these simulations techniques can give deep inside in structure and dynamics in SIB related materials helping for better design. Figure 2 shows how each neutron technique has been used in last two decades. This shows that use of neutron has an upward trend with ND as the most accessible technique. The other techniques although remain niche, are gaining ground.

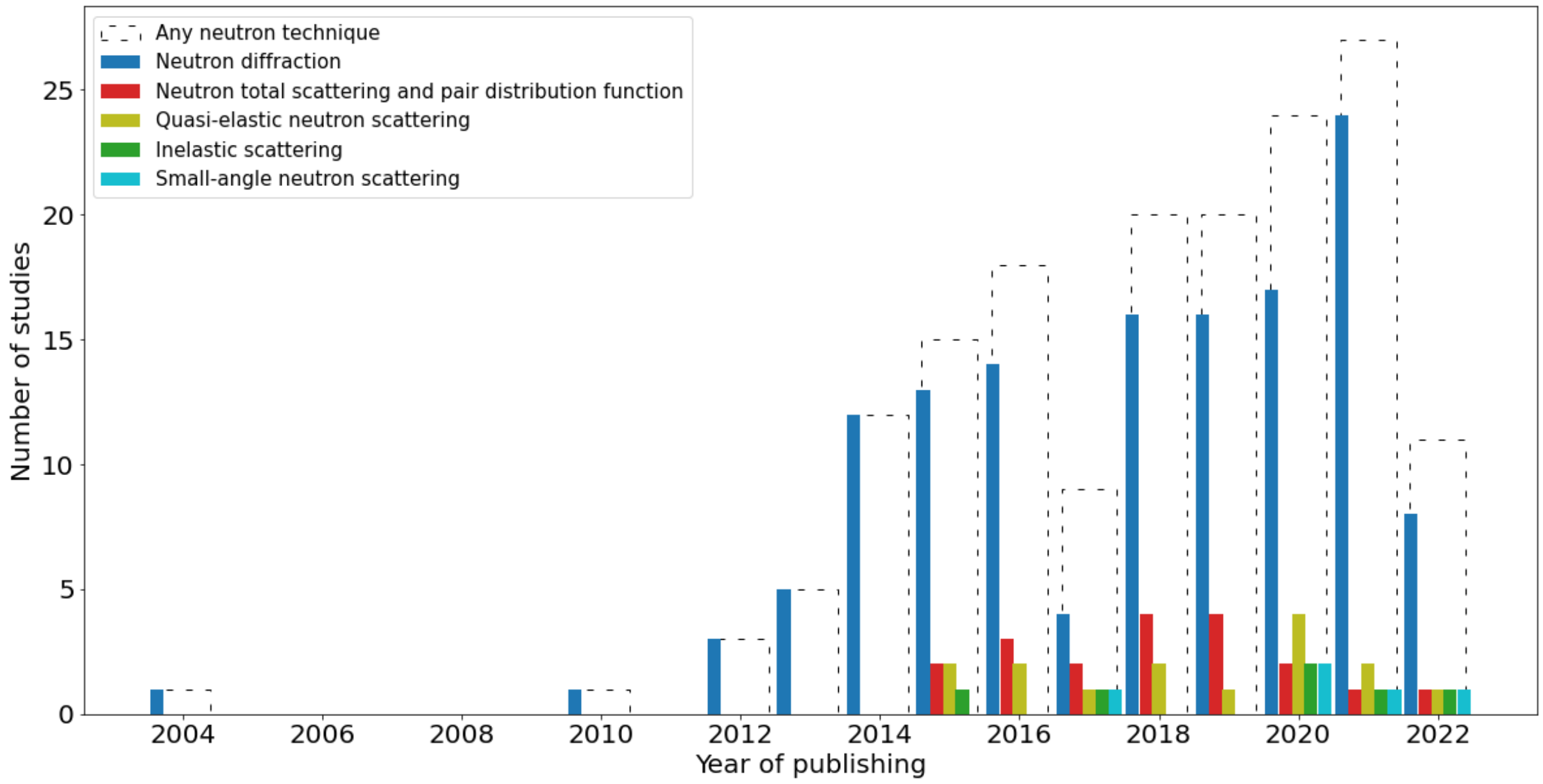

*Figure 2: The usage of neutron techniques in SIB research over the years. Only studies reviewed here are counted. The dashed bar shows the total number of SIB studies that used any neutron technique, while the coloured bars show the number of studies that used a particular technique. Therefore, each study is counted exactly once in the dashed bar but might appear in multiple coloured bars if it used multiple neutron techniques.*

## 2.1. Cathode materials

### 2.1.1. Layered oxides

Layered oxides with formula $Na_xMO_2$, where M is a transition metal or a combination of metals, consists of $MO_2$ layers formed of edge-sharing $MO_6$ octahedra, sandwiching a number of Na layers. Delmas *et al.*[27] classified these materials into multiple types based on the coordination environment of sodium ions and the occupancy of the possible close-packed sites (A, B, or C) with oxygen atoms. The crystalline group is represented in their nomenclature by a letter, P for prismatic or O for orthorhombic, etc., and the co-ordination number by a number, which is also equal to the number of transition metal layers in a repeating unit. Potential candidates for cathodes in SIB are the P2, P3, and O3-type layered oxides. Other types are either of no interest related to SIBs, or are transitionary phases in the cycling of the aforementioned three.

*Table 1: The oxygen stacking in the four phases most important in SIBs.*

| Type of lattice | Oxygen stacking |
| --- | --- |
| P2 | ABBA |
| P3 | ABBCCA |
| O3 | ABC |

The P2 and P3 type layered oxides have sodium ions in a prismatic environment with two and three transition metal layers respectively. Example structures are presented in 3. The consequence of the stacking difference (Table 1) is that the P2 phase has two distinct Na environments while in P3 phases all sodium ions are in the same environment. The two environments of the P2 type are prisms that share edges and ones that share faces with $MO_6$ octahedra, where the former is occupied preferentially. In P3 type, all prisms share one face with one octahedron and three edges with three other octahedra.[27] However, in both cases, the prismatic sodium sites share rectangular faces, allowing for direct and fast sodium ion diffusion which results in high rate performance.[28]

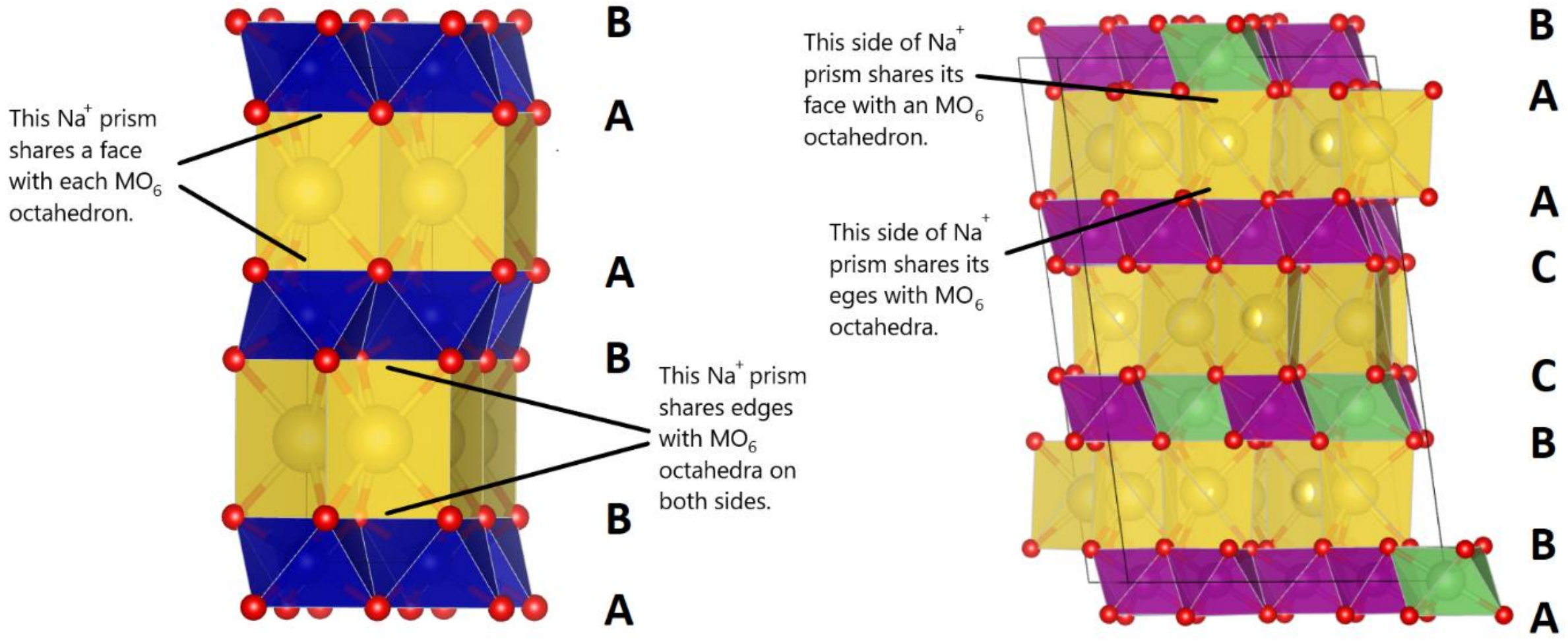

*Figure 3: Structure of sodium layered oxides from experimental data. Left: P2-type layered oxide $Na_{0.79}CoO_2$ as obtained by Beck et al.[29] Right: P3-type layered oxide $Na_{0.6}Li_{0.2}Mn_{0.8}O_2$ as obtained by Gao et al.[30]*

The O3 phase (Figure 4) contains sodium ions in an octahedral environment. It consists of $NaO_6$ octahedra sharing two faces with two $MO_6$ octahedra. The sodium octahedra do not share a face, so sodium ion diffusion has to undergo an intermediate step where the ion first enters the tetragonal gap between the octahedral sites, and only then can it diffuse to the target site. This results in much greater activation energy for diffusion, resulting in lower rate performance.[31]

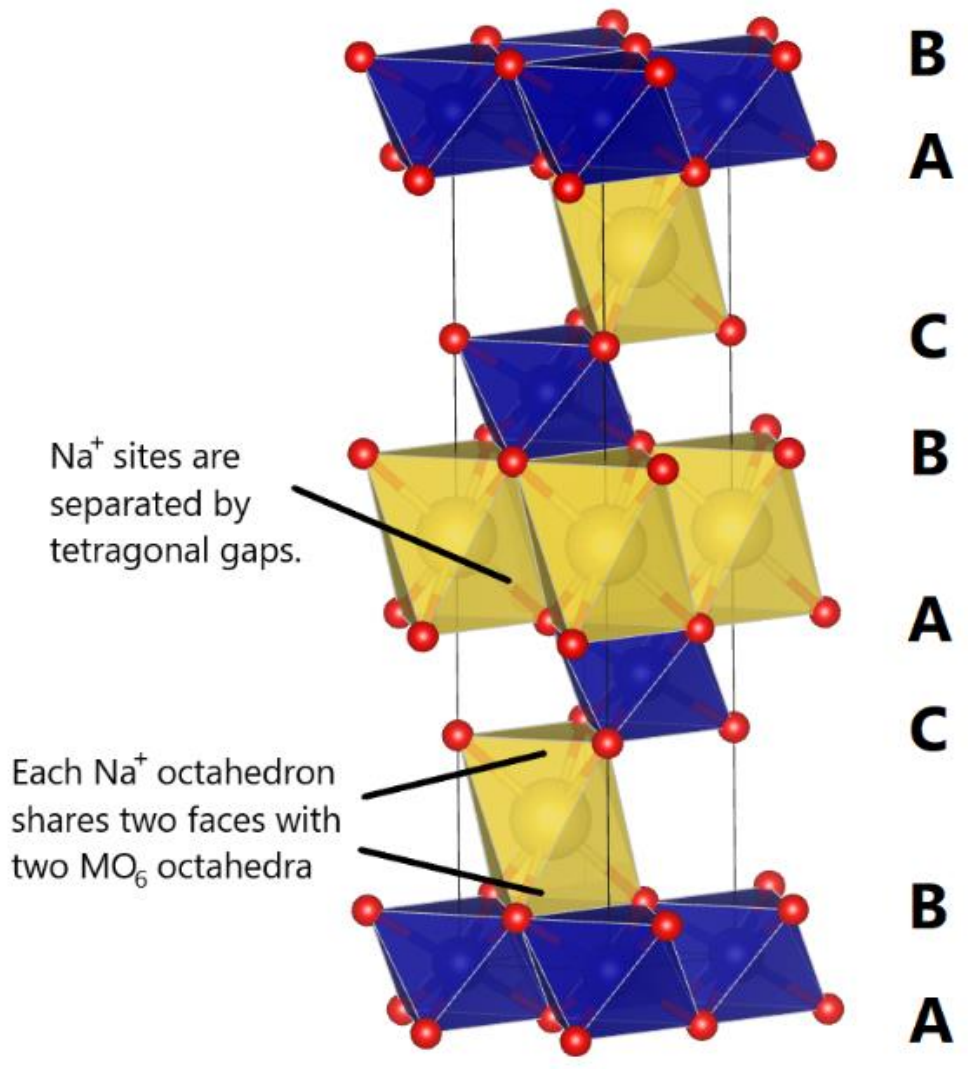

*Figure 4: Structure of an O3-type sodium layered oxide $NaCrO_2$ as obtained by Tsuchiya et al.*[32]

The P2-type phases occur in materials with lower sodium content ($0.6 < x < 0.7$) and so tend to have lower energy density. They also undergo only one phase transition, but this is often not completely reversible, leading to poor capacity retention. O3-type materials are in many ways the opposite, with high energy density due to high sodium content ($x > 0.9$). They undergo multiple reversible phase transitions and have higher cut-off voltage but are unstable in air. P3-type materials have been studied less than the other two, and so there do not seem to be any useful generalisations. Overall, layered oxides are excellent cathode materials, but great improvements are possible and necessary, whether that be through optimising synthesis methods, material morphology, or chemical composition, or through combining phases into multiphase compounds.[16,22,23,33,34]

Unfortunately, it has been pointed out that this improvement has been so far built upon empirical evidence rather than theoretical understanding.[16,22,33] leading to material discovery as mostly a luck.[13] To complicate the issue further a staggering amount of combinations of metals is possible in layered oxides.[16,22] To overcome this challenge establishing correlations between structural and electrochemical properties is the first step to efficient SIB material discovery.

The layered oxide SIB cathode materials investigated using neutrons have been listed in Table 2. Although a variety of neutron techniques have been employed in these investigations, the focus of these studies are mostly to understand specific materials, not any structure properties correlations mentioned above. Regardless, though, this as well as neutron data from studies on sodium layered oxides for purposes other than SIBs, such as on superconducting properties of $Na_xCoO_2$[35–40] or on magnetic properties of $NaMnO_2$,[41] may provide useful information to understand these materials.

As mentioned above, among all employed neutron techniques, ND is the most common. It is usually used to complement XRD to get information inaccessible to it, mainly distinguishing neighbouring elements and observing light elements. The former is often of great importance since the transitions metals employed in these layered oxides are almost exclusively from period 4 in the periodic table, making neighbours, such as Mn and Fe,[42–46] fairly common. Further, the ordering of these elements can have effects on electrochemical properties, meaning the distinction between them important. The latter issue usually manifests in materials that incorporate Li.[43,47–52] Of course, there are other specific reasons to employ ND,[53–55] and oftentimes it is used as a part of a routine.[56–60] Even then, though, the complementarity of the techniques can still provide useful information even if it was not intended. For example, Rong *et al*.[61] refined XRD pattern with $P6_3/mmc$ space group but found there were additional weak peaks, so they theorised that these could be due to the existence of a superlattice. ND provided evidence for this fact since the peaks were absent in it due to the small scattering length difference between Li and Mn. Some studies[62,63], particularly by Medarde *et al*.,[62] just show how much information can be extracted from ND experiments.

Apart from investigating the initial structure of layered oxides, understanding the structural changes during electrochemical cycling is crucially important. In situ and operando, as well as ex situ to a degree, techniques are the best and most important ways to obtain this data.[16,17,24,25] Some consider that *in situ* XRD and ND are becoming routine,[17] and indeed many of the studies in Table 2 use *in situ* XRD, however, *in situ* ND is still very rare due to its complexity. One such experiment was performed by Liu *et al*.[59] because their *in situ* XRD experiment did not explore the entire reaction due to lacking pressure in the cell. Given that the ND took 62 hours for one charge-discharge cycle, as well as the other aforementioned issues, explains why these types of experiments are rare. Nevertheless, thanks to ND's advantages, they were able to identify the structure of Z-phase, a high-voltage phase that could not be resolved with *in* situ XRD, as an O3s phase. This shows that *in situ* ND has the potential to reveal vital information difficult to access via other techniques. Thus, it is advisable that it is adopted for other critical questions, and once instrumentation and cell technology improve, it will hopefully become an indispensable tool for ND used alongside *in situ* XRD.

*Operando* ND is very rarely used method. Despite being considered the superior operando diffraction method,[16,19] there is only one study that uses it. Chen *et al*.[64] performed it to gain further evidence for their theories on the causes of capacity fading in the P2-$Na_{0.67}Fe_{0.33}Mn_{0.67}O_2$. *Operando* ND was used most likely used to distinguish between Fe and Mn, which was important because the two metals behave differently during charge-discharge cycling. Unfortunately, the importance of this successful usage of *operando* ND is undermined by the lack of details on its experimental setup. This complicates the reproducibility of the experiment, but perhaps more importantly, the application of their method to other materials.

While ND and XRD work excellent with layered oxides due to their regular, crystalline structure, not all parts of a layered oxide are ordered in all cases. The distribution of various dopants or sodium ions can play an important role in electrochemical properties of the material,[54,58,65] but it can be disordered or only partially ordered and so not appear in ND and XRD spectra. nPDF can be used to probe the local structure as well as provide some long-range ordering information, though the latter was of lesser interest in the reviewed studies since it is likely to appear in diffraction.[29,61,66–71] Instead, the local structure can be used to investigate oxygen redox thanks to nPDF's sensitivity to oxygen,[61,69] the M-O bond behaviour,[69,70] cation ordering,[29,66–68] and other properties.[61,69–72] If it is used in conjunction with other local techniques like ssNMR and xPDF to complement ND and XRD, a comprehensive structural elucidation becomes possible, something that is lacking in some of the studies which acknowledge an anomaly in their diffraction spectrum without further experimental investigation.[59,63,65]

Lastly, QENS's ability to investigate diffusion mechanism and other dynamics is unmatched, even if some studies are able to predict diffusion pathways and similar information using other techniques.[62] While computational techniques can predict dynamics information,[43,73] this should ideally be confirmed with QENS studies so that any theory stands on firm experimental grounds. Of course, this may become less important once a model in agreement with experimental observations is established, but the three QENS studies below on layered oxides are hardly enough to base any model on. The first step could be to establish a link between the different QENS results of Juranyi *et al.*[74] for $Na_{0.7}CoO_2$ at T > 400 K and those of Willis *et al.*[73] for $Na_{0.8}CoO_2$ at T = 300 K. Following that, investigate other P2-type oxides to establish the effect of various transition metal compositions on diffusion, using Chen *et al.*'s[75] work on $Na_{0.67}Ti_{0.67}Ni_{0.33}O_2$. Afterwards, O3- and P3-type materials can be studied, with final application on multiphase oxides.

*Table 2: List of studies that use neutrons to investigate layered oxide materials for their use as SIB cathode. Used abbreviations are as follows: ND = neutron diffraction, NPD = neutron powder diffraction, QENS = quasielastic neutron scattering, PDOS = phonon density of states, NTS = neutron total scattering, XRD = X-ray diffraction, PXRD = powder X-ray diffraction, SXRD = synchrotron X-ray diffraction, XANES = X-ray absorption near edge spectroscopy, XPS = X-ray photoelectron spectroscopy, EDXS = energy dispersive X-ray spectroscopy, XTS = X-ray total scattering, XAFS = X-ray absorption fine structure spectroscopy, EXAFS = extended X-ray absorption fine structure spectrometry, TXM = transmission X-ray microscopy, XRS = X-ray Raman spectroscopy, XFM = X-ray fluorescence microscopy, SXFM = synchrotron X-ray fluorescence microscopy, PDF = pair distribution function, FTIR = Fourier-transformed infrared spectroscopy, MC = Monte Carlo simulation, DFT = density functional theory simulation, AIMD = ab initio molecular dynamics simulation, CV = cyclic voltammetry, EC = electrochemical cycling, GITT = galvanostatic intermittent titration technique, PITT = potentiostatic intermittent titration technique, D = diffusivity measurement, FRA = impedance data, EIS = electrochemical impedance spectroscopy, M = magnetisation, MS = magnetic susceptibility, EPR = electron paramagnetic resonance, TEM = transmission electron microscopy, STEM = scanning transmission electron microscopy, ADF-STEM = annular-dark field scanning transmission electron microscopy, HRTEM = high resolution transmission electron microscopy, FESEM = high-resolution field emission scanning electron microscopy, NBD = nanobeam diffraction, ICP-AES = inductively coupled plasma atomic emission spectroscopy, ICP-MS = inductively coupled plasma mass spectrometry, ssNMR = solid-state nuclear magnetic resonance, MAS NMR = magic-angle spinning nuclear magnetic resonance, TGA = thermogravimetric analysis, DSC = differential scanning calorimetry, OEMS = online electrochemical mass spectrometry*

| Material | Phase | Techniques | Reference |
|---|---|---|---|
| $Mn_5O_8$ | - | nPDF, xPDF, XRD, XPS, TEM, STEM, CV, EC, DFT | Shan *et al.*[76] |
| $Na_{0.7}CoO_2$ | P2 at T > 400 K, P'2 at 290 K < T < 400 K, monoclinic at T < 290 K | NPD | Medarde *et al.*[62] |
| | | QENS, elastic forward scan, inelastic forward scan, PDOS | Juranayi *et al.*[74] |
| $Na_{0.8}CoO_2$ | Fully ordered stripe at T < 290 K, partially disordered stripe at 290 K < T < 370 K, Disordered at T > 370 K | XRD, MC, QENS, DFT, AIMD | Willis *et al.*[73] |
| $Na_{0.79}CoO_2$ | P2 | | Beck *et al.*[29] |
| $Na_{0.79}Co_{0.7}Mn_{0.3}O_2$ | P2 | NPD, nPDF, ICP-AES, PXRD, CV, EC, FRA, M | |
| $Na_2Mn_3O_7$ ($Na_{4/7}Mn_{6/7}O_2$) | P—O | SXRD, HRTEM, NTS, nPDF, XAS, XANES, EXAFS, XPS, EPR, in situ XRD, EC | Song *et al.*[68] |
| | P—O | NPD, XRD, TGA, DSC, FESEM, EIS, | Saha *et al.*[77] |
| $Na_{0.61}MnO_2$ | P2/P’2 | NPD, EC, XRD, operando XRD, SEM, EDXS, XANES, ICP | Kulka *et al.*[78] |
| $Na_{0.62}MnO_2$ | P2 | ICP-AES, PXRD, NPD, SEM, EDXS, HRTEM, XPS, CV, EC | Li *et al.*[49] |
| $Na_{0.62}Ni_{0.12}Ti_{0.12}Mn_{0.76}O_2$ | P2 | | |
| $Na_{0.67}MnO_2$ | P’2/P’2 | PXRD, SPXRD, NPD, $^{23}$Na ssNMR, EC | Clément *et al.*[79] |
| $Na_{0.67}Mn_{1-y}Mg_yO_2$ ($y \in \{0.05, 0.1\}$) | P’2 | | |
| $Na_xCr_xTi_{1-x}O_2$ ($y \in \langle 1, 0.5 \rangle$) | various, depending on synthesis temperature and composition | ND, XRD, XAS, EC | Tsuchiya *et al.*[32] |
| $Na_{0.67}Ni_{0.67}Te_{0.33}O_2$ | P2 | NPD, XRD, EIS | Bera and Yusuf[80] |
| $Na_{0.67}Ti_{0.67}Ni_{0.33}O_2$ | P2 | NPD, atomistic simulations | Shanmugam *et al.*[63] |
| $Na_{0.67}Ti_{0.67}Ni_{0.33}O_2$ | P2 | QENS, DFT | Chen *et al.*[75] |
| $Na_{0.6}Ti_{0.4}Cr_{0.6}O_2$ | P2 | NPD, MEM, PXRD, in situ XRD, ex situ XAS, TEM, EC, ICP-AES, FPMD | Wang *et al.*[81] |
| $Na_{0.67}Al_{0.1}Mn_{0.9}O_2$ | P2 | NPD, PXRD, ICP-AES, SEM, TEM, EDXS, $^{23}$Na MAS NMR, XPS, in situ XRD, EC, GITT | Xiao *et al.*[82] |
| $Na_{0.67}Fe_{0.33}Mn_{0.67}O_2$ | P2 | PXRD, XAS, EC, SEM | Chen *et al.*[64] |
| $Na_{0.67}Mn_{0.6}Fe_{0.4}O_2$ | P2 | NPD, XRD, in situ SXRD, operando SXRD, EC | Dose *et al.*[83] |
| $Na_{0.67}Mn_{0.8}Fe_{0.2}O_2$ | P2 | | |
| $Na_{0.67}Mn_{0.9}Fe_{0.1}O_2$ | P2 | | |
| $Na_{0.67}Mn_{0.5}Fe_{0.5}O_2$ | P2 | NPD, PXRD, ex situ XRD, in situ XRD, XAFS, SEM, HRTEM, XPS, ex situ EPR, DSC, EIS, EC, CV | Kong *et al.*[42] |
| | P2 | NPD, SPXRD, in situ PXRD, ICP-AES, EC, GITT | Boisse *et al.*[45] |
| | P2 | NPD, nPDF, XRD, operando XRD, operando XAS, ex situ XAS, EXAFS, SEM, EDS, ICP-AES | Wang *et al.*[72] |
| | P2 | NPD, PXRD, in situ XRD, ex situ XRD, ex situ XAS, $^{7}$Li ssNMR, SEM, HRTEM, ADF-STEM, XPS, EPR, Raman, in situ differential electrochemical MS, ex situ resonant inelastic X-ray scattering, EC, CV, GITT, DFT | Li *et al.*[51] |
| $Na_{0.67}Mn_{0.75}Fe_{0.25}O_2$<br>$Na_{0.67}Mn_{0.73}Ca_{0.02}Fe_{0.25}O_{1.98}F_{0.02}$<br>$Na_{0.66}Mg_{0.01}Mn_{0.75}Fe_{0.25}O_{1.99}F_{0.01}$<br>$Na_{0.66}Mg_{0.01}Mn_{0.73}Ca_{0.02}Fe_{0.25}O_{1,98}F_{0.02}$ | P2<br>P2<br>P2<br>P2 | NPD, XRD, in situ XRD, ex situ XRD, SEM, EDXS, ICP-AES, XPS, ex situ XPS, EC, CV, PITT | Li *et al.*[84] |
| $Na_{0.62}Mn_{0.75}Ni_{0.25}O_2$<br>$Na_{0.67}Mn_{0.67}Ni_{0.33}O_2$ | P2<br>P2 | SXRD, in situ SXRD, NPD, XANES, ICP-AES, EC | Gutierrez *et al.*[85] |
| $Na_{0.64}Mn_{0.5}Ni_{0.5}O_2$ | P2 | ND, XRD, operando XRD, ICP-AES, SEM, in situ gas chromatography-MS, XPS, EC, GITT | Li *et al.*[86] |
| $Na_{0.91}Mn_{0.5}Ni_{0.5}O_2$<br>$Na_{1.1}Mn_{0.5}Ni_{0.5}O_2$ | O3 | | |
| $Na_{0.67}Ni_{0.33}Mn_{0.67}O_{2-x}F_x$ (y $\in \{0, 0.03, 0.05, 0.07\}$) | P2 | NPD, PXRD, operando XRD, EDXS, XPS, Raman, $^{19}$F MAS NMR, TEM, HRTEM, STEM, XAFS, EC, CV, GITT | Liu *et al.*[50] |
| $Na_{0.67}Ni_{0.33}Mn_{0.67}O_2$<br>$Na_{0.67}Ni_{0.167}Mn_{0.5}Fe_{0.33}O_2$ | P2<br>P2 | SXRD, operando XRD, NPD, ADF-STEM, $^{57}$Fe Mossbauer | Somerville *et al.*[87] |
| $Na_{0.67}Mn_{0.67}Ni_{0.33}O_2$<br>$Na_{0.67}Li_{0.11}[Li_{0.11}Mn_{0.67}Ni_{0.22}]O_2$ | P2<br>P2 | NPD, PXRD, in situ XRD, SEM, EDXS, HRTEM, XPS, EC, EIS, ICP-AES | Li *et al.*[48] |
| $Na_{5/6}Li_{1/4}Mn_{3/4}O_2$ | P2 | ICP, XRD, SXRD, ND, SEM, EC | Yabuuchi *et al.*[88] |
| $Na_{0.72}Li_{0.24}Mn_{0.75}O_2$ | P2 | NPD, nPDF, EC, XRD, in situ XRD, SEM, XAS, XANES, STEM | Rong *et al.*[89] |
| $Na_{0.67}Li_{0.22}Mn_{0.78}O_2$<br>$Na_{0.75}Li_{0.15}Ni_{0.15}Mn_{0.7}O_2$ | P2<br>P2 | NPD, XRD, operando XRD, SEM, FE-TEM, XAS, $^{7}$Li NMR, ICP-AES, secondary ion MS, EC, GITT, operando differential electrochemical MS, DFT | Voronina *et al.*[52] |
| $Na_{0.6}Li_{0.2-x}Al_xMn_{0.8}O_2$ ($0 \le x \le 4$) | P2 | NPD, PXRD, operando XRD, secondary ion MS, XPS, XAS, EC, HR-TEM, TEM, FE-SEM, XANES, DFT | Yoon *et al.*[90] |
| $Na_{0.67}Mn_{0.75}Ni_{0.25}O_2$<br>$Na_{0.67}Li_{0.125}Mn_{0.625}Ni_{0.25}O_2$ | P2<br>P2 | NPD, PXRD, operando SXRD, ICP-AES, XPS, ex situ XAS, EC, GITT, CV, DFT | Park *et al.*[56] |
| $Na_{0.8}Li_{0.12}Ni_{0.22}Mn_{0.66}O_2$<br>$Na_{0.83}Li_{0.07}Ni_{0.31}Mn_{0.62}O_2$ | P2<br>P2 | ICP-AES, NPD, $^{7}$Li ssNMR, in situ SXRD, XAS, EC | Xu *et al.*[91] |
| $Na_{0.66}Li_{0.18}Fe_{0.12}Mn_{0.7}O_2$ | P2 | XRD, SXRD, operando XRD, operando SXRD, FESEM, XANES, ND, $^{7}$Li MAS NMR, EC | Yang *et al.*[92] |
| $Na_{0.67}Li_{0.2}Mn_{0.8}O_2$<br>$Na_{0.67}Li_{0.2}Fe_{0.2}Mn_{0.6}O_2$<br>$Na_{0.67}Li_{0.2}Fe_xMn_{0.8-x}O_2$ ($x \in \{0.4, 0.6\}$)<br>$Na_{0.67}Li_{0.2}Fe_{0.8}O_2$<br>$Na_{1-x}Li_xFe_{0.5}Mn_{0.5}O_2$ ($x \in \{0, 0.1, 0.4\}$) | P2<br>P2/O3<br>O3<br>Not layered<br>O3 | PXRD, SPXRD, NPD, ex situ ICP-AES, ex situ XAFS, in situ SXRD, EC, DFT | Wang *et al.*[93] |
| $Na_{0.67}Li_{0.18}Mn_{0.8}Fe_{0.2}O_2$ | P2/O3 | NPD, operando SXRD, EC, ICP-AES | Stansby *et al.*[65] |
| $Na_{0.67}Li_{0.2}Mn_{0.6}Fe_{0.2}O_2$ | P2/O3 | NPD, SPXRD, ICP-AES, in situ SXRD, EC, DFT | Wang *et al.*[58] |
| $Na_{0.67}MnO_2$<br>$Na_{0.67}M_{0.1}Mn_{0.9}O_2$ ($M \in \{Li, Mg, Al\}$)<br>$Na_{0.67}Al_{0.1}Fe_{0.05}Mn_{0.85}O_2$ | P2<br>P2<br>P2 | NPD, PXRD, in situ SXRD, SEM, TEM, ICP-AES, $^{23}$Na MAS NMR, EC, GITT | Liu *et al.*[54] |
| $Na_{0.67}Mn_{0.5+y}Ni_yFe_{0.5-2y}O_2$ (y ∈ {0, 0.1, 0.15}) | P2 | PXRD, NPD, operando XRD, xPDF, Mossbauer spectrometry, ICP-AES, EDXS, EC | Talaie *et al.*[94] |
| $Na_{0.67}Mn_{0.5+y}Ni_yFe_{0.5-2y}O_2$ (y ∈ {0, 0.1, 0.15}) | P2 | PXRD, NPD, SEM, EDXS, ICP-AES, TGA, EC, OEMS | Duffort *et al.*[95] |
| $Na_{0.67}Ni_{0.33-x}Mg_xMn_{0.67}O_2$ (x ∈ ⟨0, 0.2⟩) | P2 | PXRD, NPD, XANES, operando SXRD, ICP-MS, EC, DFT, AIMD | Tapia-Ruiz *et al.*[96] |
| $Na_{0.67}Mn_{0.8}Fe_{0.1}Ti_{0.1}O_2$ | P2 | PXRD, in situ XRD, NPD, ICP-AES, SEM, EDXS, $^{23}$Na MAS NMR, Mossbauer, EC | Han *et al.*[97] |
| $Na_{0.67}Mn_{0.8}Fe_{0.1}Ti_{0.1}O_2$ | P2 | NPD, in situ SXRD, operando SXRD, $^{23}$Na MAS NMR, EC, CV | Han *et al.*[98] |
| $Na_{0.67}Mn_{0.8}Fe_{0.1}Ti_{0.1}O_2$ | P2 | NPD, SPXRD, operando XRD, in situ SXRD | Goonetilleke *et al.*[99] |
| $Na_{0.67}Mn_{0.9-x}Ni_xFe_{0.05}Ti_{0.05}O_2$ (x ∈ {0.1, 0.2}) | P2 | NPD, XRD, TEM, ICP-AES, $^{23}$Na MAS NMR, ex situ XRD, in situ SXRD, EC | Ortiz-Vitoriano *et al.*[44] |
| $Na_{0.67}Mg_{0.11}Ti_{0.11}Mn_{0.56}Ni_{0.22}O_2$ | P2 | NPD, PXRD, in situ XRD, ex situ XRD, HRTEM, EDXS, SEM, XPS, EC, EIS | Li *et al.*[100] |
| $Na_{0.7}Mg_xMn_{0.8}Ni_{0.1}Co_{0.1}O_2$ (x ∈ {0, 0.05}) | P2 | PXRD, NPD, SEM, XPS, EC | Li *et al.*[47] |
| $Na_{0.7}Mn_{0.75}Fe_{0.75}O_2$<br>$Na_{0.7}Mn_{0.75}Fe_{0.15}Ni_{0.1}O_2$<br>$Na_{0.7}Mn_{0.75}Fe_{0.15}Co_{0.1}O_2$<br>$Na_{0.7}Mn_{0.75}Fe_{0.15}Ni_{0.05}Co_{0.05}O_2$ | P2<br>P2<br>P2<br>P2 | PXRD, NPD, SEM, XPS, EC, CV | Wang *et al.*[101] |
| Na-$_{0.75}Co_{0.125}Cu_{0.125}Fe_{0.125}Ni_{0.125}Mn_{0.5}O_2$ | P2/P3 | NPD, SXRD, XAFS, EXAFS, FESEM, SXFM, HRTEM, ICP-AES, EC | Rahman *et al.*[60] |
| $Na_{0.6}Li_{0.2}Mn_{0.8}O_2$ | P3 | EC, XRD, SEM, ICP-AES, in situ XRD, XPS, SXRD, xPDF, NPD, nPDF, ex situ hXAS, ex situ sXAS, TXM, TXM XANES | Rong *et al.*[102] |
| $Na_{0.67}Mg_{0.33}Mn_{0.67}O_2$ | P3 | SXRD, xPDF, nPDF, XAS, XANES, EXAFS, EC | Song *et al.*[67] |
| $Na_{0.5}Ni_{0.25}Mn_{0.75}O_2$ | P3 | NPD, in situ NPD, PXRD, in situ XRD, DFT, ex situ XAS, EC, GITT, ICP-AES | Liu *et al.*[59] |
| $Fe_xM_{1-x}O_2$ (M ∈ {Mn, Co, Ni}; x ∈ {0.25, 0.5, 0.75}) | rippling | DFT | Chen *et al.*[103] |
| $Na_{0.111}Fe_xM_{1-x}O_2$ (M ∈ {Mn, Co, Ni}; x ∈ {0.25, 0.5, 0.75}) | rippling | DFT | |
| $NaMn_{0.5}Fe_{0.5}O_2$ | O3 | NPD, DFT | |
| $NaFe_{0.5}Co_{0.5}O_2$ | O3 | NPD, in situ XRD, DFT | |
| $NaMn_{0.25}Fe_{0.25}Ni_{0.25}Co_{0.25}O_2$ | O3 | NPD, in situ XRD, DFT | |
| $NaLi_{0.33}Ir_{0.67}O_2$ | O3 | NPD, XRD, SXRD, in situ XRD, EC, DFT | Perez *et al.*[55] |
| $Na_2IrO_3$ (ie. $Na[Na_{0.33}Ir_{0.67}]O_2$) | O3 | NPD, in situ NPD, XRD, ex situ XRD, in situ XRD, SXRD, XPS, TEM, electron diffraction, high angle ADF-STEM, annular bright field STEM, EDXS, OEMS, EC, DFT | Perez *et al.*[104] |
| $NaLi_{0.32}Mn_{0.68}O_2$ | O3 | NPD, XRD, TEM, $^{6}Li$ MAS NMR, $^{23}Na$ MAS NMR, GITT, EC, HAXPES, mRIXS, ex situ XAS, operando XAS, MS, DFT | Wang *et al.*[105] |
| $NaNi_{0.67}Sb_{0.33}O_2$ | O3 | PXRD, SXRD, NPD, nPDF, EC, TEM, $^{23}Na$ ssNMR | Ma *et al.*[66] |
| $Na_{0.625}MnO_2$ | O’3 | In-situ XRD, STEM, SXRD, NPD, MS, DFT | Li *et al.*[106] |
| $Na_{0.9}MnO_2$ | O3 | ICP-AES, ssNMR, XANES, XPS, NPD | Dally *et al.*[107] |
| $NaFe_{1-y}Co_yO_2$ (y ∈ {0, 0.4, 0.5, 0.6, 1}) | O3 | ND, XRD, operando XRD, SXRD, XAS, ICP-AES, EC, MC, DFT | Kubota *et al.*[108] |
| $Na_{0.77}Mn_{0.33}Fe_{0.67}O_2$ | O3 | NPD, SPXRD, in situ PXRD, ex situ SPXRD, in situ XANES, in situ | Boisse *et al.*[109] |
| | | $^{57}$Fe Mossbauer, ICP-AES, EC, GITT | |
| $NaNi_{0.28}Mn_{0.23}Fe_{0.5}O_2$<br>NCG- $NaNi_{0.28}Mn_{0.23}Fe_{0.5}O_2$<br>NCG- $NaNi_{0.28}Mn_{0.23}Fe_{0.5}O_2$-Y | O3<br>O3<br>O3 | ND, XRD, SEM, TEM, XPS, ICP-AES, EC, PITT, CV, EIS, DSC, DFT | Wu *et al.*[43] |
| $NaNi_{0.33}Mn_{0.33}Fe_{0.33}O_2$<br>$Na_3Zr_2Si_2PO_{12}$- $NaNi_{0.33}Mn_{0.33}Fe_{0.33}O_2$ | O3 | ND, XRD, SXRD, XPS, DSC, ICP, SEM, HAADF-STEM, ABF-STEM, EC, CV, EIS, PITT, DFT, AIMD | Li *et al.*[46] |
| $NaNi_{0.3}Fe_{0.4}Mn_{0.3}O_2$<br>$Na_{0.95}Li_{0.05}Ni_{0.3}Fe_{0.4}Mn_{0.3}O_2$<br>$NaNi_{0.3}Fe_{0.4}Mn_{0.3}O_{1.95}F_{0.05}$<br>$Na_{0.9}Li_{0.05}Ni_{0.3}Fe_{0.4}Mn_{0.3}O_{1.95}F_{0.05}$ | O3<br>O3<br>O3<br>O3 | NPD, XRD, EDXS, ICP-AES, FTIR, SEM, HRTEM, NBD, EC | You *et al.*[110] |
| $Na_xMn_{0.33}Fe_{0.33}Cu_{0.167}Mg_{0.167}O_2$ ($x \in \{1, 0.9, 0.8, 0.7\}$) | O3 | PXRD, NPD, XAS, SEM, XPS, DSC, ICP-AES, EC, EIS | Li *et al.*[111] |
| $Na_{0.89}Li_{0.037}Cu_{0.189}Fe_{0.274}Mn_{0.499}O_2$ | O3 | NPD, PXRD, XPS, XAS, XRS, FESEM, EDXS, ICP-AES, TXM, EC | Rahman *et al.*[112] |
| $NaTi_{0.25}Fe_{0.25}Co_{0.25}Ni_{0.25}O_2$ | O3 | NPD, nPDF, XAS, XANES, EXAFS, in situ XRD, TEM, HRSTEM, DFT, EC | Kim *et al.*[70] |
| $Na_{0.9}Ti_{0.55}Ni_{0.45}O_2$<br>$Na_{0.9}Ti_{0.5}Mn_{0.05}Co_{0.15}Ni_{0.3}O_2$ | O3<br>O3 | NPD, XRD, SEM, TEM, XPS, in situ XRD, ex situ XPS, EIS, EC, DFT, ICP-AES | Jiang *et al.*[113] |
| $Na_{0.94}Ti_{0.1}Mn_{0.43}Ni_{0.47}O_2$<br>$Na_{0.94}Ti_{0.15}Mn_{0.3}Fe_{0.16}Ni_{0.29}Cu_{0.1}O_2$ | O3<br>O3 | NPD, XRD, SEM, in situ XRD, EC, CV, ICP-AES | Guo *et al.*[53] |
| $Na_xMnO_2$ | O’3 | In-situ SXRD, in-situ XRD, HRTEM, STEM, NPD, MS, DFT | Chen *et al.*[114] |
| $Na_{0.27}MnO_2.0.63H_2O$ | birnessite | ICP-MS, EDXS, TGA, TEM, CV, D, OCV, EIS, in situ XRD, XTS, NTS, PDF, DFT | Shan *et al.*[71] |
| $Na_{0.36}MnO_2.0.2H_2O$ | birnessite | NPD, PXRD, TEM, HRTEM, EDXS, ADF-STEM, TGA, differential thermal analysis, MS | Bakaimi *et al.*[115] |
| $Na_{0.653}Mn_{0.929}O_2$ | disordered | NPD*, PXRD, SEM, EDXS, XPS, EC, CV, DFT | Zhao *et al.*[116] |

### 2.1.2. Polyanionic and NASICON-type materials

Polyanionic compounds form another set of highly researched and promising SIB cathode materials. They normally consist of a single type of polyanion $(XO_4)^{n-}$, though mixed polyanionic compounds are not rare. Mixed polyaonic compounds which may also contain other polyanions (eg. $(PO_4)^{n-}$ and $(SO_4)^{n-}$), or multiple structural units of the same polyanion (eg. $(PO_4)^{n-}$ and $P_2O_7$) in addition to single anions (eg. $(PO_4)^{n-}$ and $F^-$), have similar yet often better structural properties. In either case, the polyanions form stable frameworks with strong covalent bonds which provide these compounds with their main advantages over layered oxides; high structural and thermal stability, few phase transitions, and little to no volume changes; which together ensure stable structures during charge-discharge cycles and safe operations. High redox potentials can also be achieved. However, they have low ionic conductivity, resulting in the necessity for special treatments. This could be avoided by using NASICON-type (Na superionic conductor) materials, compounds with general formula $Na_xMM'(XO_4)_3$ and structurally similar to sodium super ionic conductors known to have excellent conductivity, but many such materials tested for use in SIBs have conductivities similar to other polyanionic compounds.[16,22,117–120]

The NASICON-type materials composed of $PO_4$, $SiO_4$, and $ZrO_6$ polyhedral units forming a zigzag framework. They crystallize in either a monoclinic (C2/c) or a rhombohedral ($R\bar{3}c$) crystallographic structure depending on their compositions and ambient temperatures. A structural transformation may occur above room temperature and sometime well below operating temperatures. For example, the stable monoclinic phase of standard NASICON (i.e., $Na_3Zr_2Si_2PO_{12}$ stoichiometry) at room temperature transforms to the rhombohedral phase around 430 K.

Therefore, the discovery of highly conductive materials is of great importance accompanying improvement of other properties such as storage capacity, energy density, structural and dynamical stability at operating condition. Achieving these by using cheaper transition elements, such as iron and manganese instead of the so far most successful vanadium, or alkaline dopants, such as Mg or Ba, is the current focus of investigations.[16,22,117–120]

Investigations using neutrons are increasing for realising these goals as listed in Table 3. All of those reported results have used ND to probe structure of these materials. There are some studies that investigated deeper by performing calculations on diffraction data. Kan *et al.*[137] used maps of low bond valence mismatch derived from ND and XRD data to visualise diffusion pathways. On the other hand Nishimura *et al.*[131] used maximum entropy method (MEM) to calculate neutron scattering length density distributions from ND data to determine the diffusion pathways. However, no attempt has been made to use any neutron spectroscopy techniques or QENS to directly probe the diffusion in these materials and through that come closer to understanding the all-important issue of low conductivity.

None of the ND experiments were performed *in situ* or *operando*, but this is less surprising than in the case of layered oxides since phase changes are rare and thus do not affect the electrochemical properties of most polyanionic compounds.[118] Still, though, *in situ* and *operando* XRD are of interest, with 4 of the studies using such,[121,123,125,135] so it is likely that the neutron equivalent can provide valuable information.

As for other techniques, nPDF is rarely necessary, given the crystallinity of polyanionic compounds, but there are cases where it could have been of use. For example, He *et al.*[132] attributed the observed abnormal electrode behaviour and higher charge capacity to proton impurities invisible to ND and XRD, possibly due to their random distribution. SANS, on the other hand, could become invaluable in the future. The research interest in the polyanionic cathode field is turning to nanoparticles and coating for the solution to low conductivity, for study of which SANS is well equipped.

*Table 3: List of papers that used neutrons to research a polyanionic NIB cathode material. Most abbreviations used can be found in Table 2, the new ones are as follows: SAED = selected-area electron diffraction.*

| Type | Formula | Techniques used | Reference |
|---|---|---|---|
| NASICON | $Na_3V_2(PO_4)_3$ | NPD, XRD, Raman, FESEM, MS | Kao *et al.*[128] |
| | $Na_4MnCr(PO_4)_3$ | ND, SXRD, in situ XRD, ex situ SXRD, SEM, TEM, TGA, XPS, ex situ XANES, EC, CV, EIS, GITT, DFT | Zhang *et al.*[123] |
| | $Na_xMnV(PO_4)_3$ | NPD, XRD, in situ XRD, ICP-AES, SEM, TEM, XAS, XPS, $^{23}$Na ssNMR, EC, GITT, CV, DFT | Hou *et al.*[121] |
| Polyanionic compounds | $NaMnFe_2(PO_4)_3$ | ND, XRD, SEM, ICP-AES, $^{57}$Fe Mossbauer spectroscopy, DFT, EC | Trad *et al.*[148] |
| | $Na_3V(PO_4)_2$ | NPD, XRD, ICP-MS, XANES, $^{51}$V ssNMR, ex situ $^{51}$V ssNMR, EC, CV | Liu *et al.*[124] |
| | $NaFePO_4$ | NPD, XRD, MS, $^{57}$Fe Mossbauer, HC | Avdeev *et al.*[135] |
| | $Na_3Fe_3(PO_4)_4$ | NPD, SXRD, Brunauer-Emmet-Teller, SEM, TEM, MS, EC | Shinde *et al.*[126] |
| | $Na_4Ni_7(PO_4)_6$ | NPD, MS | Xia *et al.*[127] |
| | $NaCoCr_2(PO_4)_3$<br>$NaNiCr_2(PO_4)_3$<br>$Na_2Ni_2Cr(PO_4)_3$ | NPD, XRD, EDX, EC, CV | Yahia *et al.*[130] |
| | $Na_2Co_2Fe(PO_4)_3$ | NPD, XRD, EC | Essehli *et al.*[151] |
| | $Na_2CoP_2O_7$ | NPD, XRD, MS, HC | Barpanda *et al.*[138] |
| | $Na_2CoZnP_2O_7$<br>$Na_2CoFeP_2O_7$ | ND, XRD, $^{31}$P ss-NMR, DFT | Ferrara *et al.*[153] |
| | $Na_{6.88}V_{2.81}(P_2O_7)_4$ | NPD, XRD, operando XRD, ex situ XANES, ICP-AES, HRTEM, SAED, EDX, Raman, TGA, EC, CV, EIS, GITT | Li *et al.*[149] |
| | $Na_{2.5}Fe_{1.75}(SO_4)_3$ | NPD, SXRD, EC | Nishimura *et al.*[131] |
| | $Na_2M(SO_4)_2.4H_2O$ (M ∈ {Mg, Fe, Co, Ni, Zn}) $Na_2M(SO_4)_2$ (M ∈ {Fe, Co}) | NPD, XRD, ex situ XRD, SXRD, SAED, $^{57}$Fe Mossbauer, IR, TGA, DSC, EC | Reynaud *et al.*[150] |
| Mixed polyanionic compounds | $NaFe(SO_4)_{1.5}X_{0.5}$ (X ∈ {$SeO_4$; $HPO_4$; $PO_3F$}) | NPD, XRD, TGA, IR, Raman, SEM, $^{57}$Fe Mossbauer, EIS, EC | Trussov *et al.*[122] |
| | $LiNaCoPO_4F$ | NPD, XRD, EDX, MS, HC, DFT, EC | Yahia *et al.*[136] |
| | $NaVOPO_4$ | NPD, XRD, SXRD, ICP-SEM, EC, CV | He *et al.*[132] |
| | $Na_3V_2(PO_4)_2F_3$ | NPD, XRD, ex situ XRD, FESEM, ICP-AES, TGA, EC, DFT | Shakoor *et al.*[141] |
| | $Na_{3-x}V_2(PO_{4-x}F_x)_2$ (x ∈ {0, 0.1, 0.15, 0.3}) | NPD, XRD, FESEM, EDX, HRTEM, XPS, EC, CV | Muruganantham *et al.*[129] |
| | $Na_3V_2(PO_4)_2F_3$ | NPD, XRD, SXRD, SEM, ED, MS, EC, ICP-AES | Bianchini *et al.*[146] |
| | $Na_2FePO_4F$ | NPD, XRD, MS | Avdeev *et al.*[133] |
| | | NPD, XRD, TGA, $^{57}$Fe Mossbauer, Raman, IR, TEM, SAED, EELS, electron diffraction tomography, SEM, EC, PITT, EIS, CV, DFT, MC | Kirsanova *et al.*[134] |
| | $Na_{2-x}Li_xFePO_4F$ (x ∈ ⟨0, 1⟩) | ND, XRD, ex situ XRD, $^{57}$Fe Mossbauer, SEM, ex situ EDX, EC, GITT | Kosova *et al.*[142] |
| | $Na_3V_{2-y}O_{2-y}Fe_y(PO_4)_2F_{1+y}$ (y ∈ {0, 0.1, 0.2, 0.3}) | NPD, XRD, in situ SXRD, SEM, ESR, $^{23}$Na MAS-NMR, EC | Palomares *et al.*[125] |
| | $Na_2MnPO_4F$ | NPD, MS | Lochab *et al.*[152] |
| | $Na_3V_{2-x}Mn_xO_{2y}(PO_4)_2F_{3-2y}$ (x ∈ ⟨0, 1⟩ y ∈ ⟨0, 1⟩) | NPD, XRD, SEM, EELS, HAADF-STEM, EDX, TEM, EC | Iarchuk *et al.*[144] |
| | $Li_xNa_{4-x}Fe_3(PO_4)_2(P_2O_7)$ (x ∈ ⟨0, 3⟩) | NPD, XRD, in situ XRD, $^{57}$Fe Mossbauer, ICP-AES, TGA, DSC, EC, DFT | Kim *et al.*[140] |

| | | |
|---|---|---|
| $Na_xFe_3(PO_4)_2(P_2O_7)$ ($x \in \langle 1, 4 \rangle$) | NPD, ex situ XRD, in situ XRD, FESEM, TEM, XANES, XAS, DSC, ICP-AES, PITT, EC, AIMD, DFT | Kim *et al.*[139] |
| $Na_4NiP_2O_7F_2$ | ND, XRD, FESEM, EDX, TGA, EC, MD | Kundu *et al.*[147] |
| $Na_3TiP_3O_9N$ | NPD, XRD, SXRD, in situ SXRD, XAS, SEM, EDX, XANES, EC, GITT, TGA, DFT | Liu *et al.*[143] |
| $Fe_3P_5SiO_{19}$ | NPD, XRD, EDX, EC, SEM | Kan *et al.*[137] |

### 2.1.3. Other cathode materials

Prussian blue and its analogues (PBAs) form the last large family of SIB cathode materials. Their main advantage is the combination of low price and good electrochemical properties, though the latter still has to be improved, especially the capacity and charge-discharge cyclability. Doing that, however, is complicated due to how these properties are tied to the presence of vacancies and water within the structure. While the former is desirable to be removed, the presence of water has both advantages and disadvantages. As such, it might be necessary to take a holistic approach to PBA cathode development: the material, its coating or other modifications, and the electrolyte might have to be taken into account simultaneously.[16,34]

To do that, large amounts of experimental data is necessary, but only a couple of studies used a neutron technique, such ND, to study PBA for SIB cathode application.[154,155] That is in spite of the fact that some articles on the topic explicitly call for future research using neutrons.[156,157] Of course, some insights can be gained from general neutron research on PBAs[158–161] or from neutron studies on PBA applications to potassium-ion batteries,[162,163] but further extensive research into PBAs for SIBs using neutrons is highly required. For example, QENS could be used to study water diffusion in a set of PBAs intercalated with sodium.

Other than that, some other SIB cathode materials have been studied with neutrons. Wiedemann *et al.*[164] used ND to investigate an old yet little-known material $Na_xTiS_2$, and $Ti_3C_2T_x$ (where T represents oxygen, hydroxyl, and fluorine groups on material surface), Nava-Avendaño *et al.*[165] studied sodium manganese fluorides with ND, and Park *et al.*[166] used SANS to examine functionalised reduced graphene oxide.

## 2.2. Anode materials

### 2.2.1. Carbonaceous materials

Carbonaceous materials are among the most promising for anodes, thanks to their low cost and good voltage plateaus. Among them, hard carbon, a low density microporous amorphous material that cannot be turned into crystalline graphite even at high temperature, is the most popular exhibiting the best properties, except its low initial Coulombic efficiency. However, its structure is complex and varies depending on its precursor and manufacturing conditions. This has put a challenge for achieving any reliable mechanism of sodium storage in this material and more research is required.[16,18]

Neutron techniques are useful for providing information valuable in this endeavour. Particularly, neutron total scattering and nPDF are highly suited since they can probe amorphous systems to study their structure, understanding of which remains insufficient.[167] This makes them the most popular neutron technique. They have been applied to a variety of hard carbons,[168–173] a doped material,[174] as well as a solid electrolyte interphase layer on a hard carbon anode.[175] However, this is still a rarity compared to all studies that have been done on hard carbon.

Another technique that has great potential in this field is SANS, which can be used to study pores and other surface features. This is of import because it also affects electrochemical properties.[167] Its application has already proven successful in multiple studies,[168,169,176] but it has the potential to be utilised *in situ*,[177] and so aid in the next step in hard carbon anode research.[167]

Other neutron techniques, such as ND, INS and QENS can also be used to compliment results from XRD,[167] and Raman spectroscopy and to obtain crucial information on sodium ion dynamics through hard carbon. The analysis of these data is challenging due to low incoherent scattering cross section of carbon, which can be overcome by polarisation analysis and complementing with molecular dynamics simulations as explained in later section. The other important factor is the slow diffusion of ions in electrode materials. To probe this slow dynamics, very high resolution neutron spectroscopic instrument is required, which is still rare.

As carbonaceous materials other than hard carbon have shown worse properties as anode materials, there is less interest in them in general. Despite that, Jian *et al.*[178] used nPDF to investigate soft carbon, a material similar to hard carbon except that it can be turned into graphite, and Wu *et al.*[179] studied Super P carbon, a graphitic material, using ND.

### 2.2.2. Metal-based compounds

This is a broad category with a variety of properties and sodium ion storage mechanisms, where metal based compounds have been explored as anode materials. Titanium-based oxides are considered the most promising thanks to their low cost, ease of processing and non-toxicity, but oxides, sulphides, and phosphides of tin, antimony, molybdenum, and niobium are also highly regarded.[16,18] Among these, however, only titanium compounds have been examined with neutrons. A variety of sodium titanates,[180–184] including an aluminium-containing one,[185] $H_2Ti_3O_7$,[182,186] and a pyrophosphate[187] have all been studied. Mostly, just ND is used to determine structure, but Bhat *et al*.[180] also transformed the ND data into nPDF to study the change in local structure upon sodium ion intercalation in $Na_2Ti_9O_{19}$, and Mittal *et al.*[184] used QENS and INS to study sodium ion diffusion in $Na_2Ti_3O_7$. *In situ* and *operando* ND can also help understand these materials further, enabling further improvements. However, these techniques are more useful for studying titanium materials improved through doping or defects, so if future research instead focuses on nanostructured or composite fabrication, they might continue seeing limited use.

### 2.2.3. Polyanionic type and other anode materials

While not often mentioned among anode materials,[18] multiple polyanionic compounds have been studied for this purpose,[149,188–190] although one of those is for a sodium-sulphur battery and is an investigation of the entire setup.[190] All of these studies also only used ND. Additionally, a nanoscaled biphasic cobalt-manganese oxide[191] and germanium-doped $Fe_2O_3$ nanofibers[192] have also been considered as anode materials, using nPDF and ND respectively. Lastly, multiple layered oxides have been studied as both cathodes and anodes using a variety of techniques.[32,63,75,76,81]

Recently MXene type materials have received considerable interest as anode materials due to their excellent electrochemical properties and 2D layered structure.[18] Even having this popularity, limited studies have been done using neutrons, such as by Osti *et al*.[193] with INS, by Ferrara *et al*.[194] with ND, and by Brady *et al*.[195] with nPDF. Similar situation is with organic anodes, another promising group,[16,18] on which no neutron studies regarding SIB usage has been reported so far.

## 2.3. Electrolyte materials

There are many electrolyte materials that have been investigated for use in SIBs, such as liquid non-aqueous electrolytes based on carbonate esters, ethers, and ionic liquids, aqueous electrolytes, and (quasi-)solid-state electrolytes (SSE) such as polymers or inorganic solids. However, many widely used compounds show lacking performance in SIBs, and due to the necessity for electrolyte to be compatible with both cathode and anode materials, the development of new electrolyte materials and the understanding of how they interact with electrodes is of utmost importance.[16,196]

Among these, mostly solid electrolytes are studied with various neutron techniques, such as borates[198–201] and carbaborates,[199,202–205] phosphorus[206,207] and antimony[208–211] sulphides, NASICON-type materials,[212–217] and other compounds[218–225]. There are some neutron studies of aqueous electrolytes, but they are focused on the electrode material rather than electrolytes.[71,76,191] There is only one report of inorganic salt in organic solvent type electrolyte, which is the traditional LIB and SIB electrolyte materials, has been studied with neutrons. Some investigations have also been done on glymes type electrolytes.[197]

Given the crystallinity of many of these materials, ND is widely used,[201,206,216–220,222,224,207,209–215] and often it is the only neutron technique employed.[201,209,219,220,222,224,210–215,217,218] Most of these studies did not use *in situ* or *operando* in the sense that a full cell was cycled through charging and discharging process. Having said that, charging-discharging cycle is less important in the structural analysis since electrolytes, unlike electrodes, shouldn't experience significant change in density of sodium. Instead, phase changes with temperature are of greater importance, and there are few reports of studying this temperature effect in *in situ* condition,[210,211,219] MEM has also been utilised to visualise diffusion pathways[213,216,218] as well as sodium distribution,[216] and in only one study ND data was converted into total scattering factor.[221]

Unlike for other SIB materials, QENS is more common in studying electrolytes,[198,199,202–205,208,223,225] though most of them are on (carbo)boranes where the diffusion of protons is studied, thanks to the large neutron cross-section of hydrogen.[198,199,202–205] Apart from performing QENS, diffusion of protons has been studied using elastic forward scans[199,203] and neutron vibrational spectroscopies,[199,200,202,203] As for other techniques, partial nPDF and inelastic neutron scattering are also performed to investigate medium range structures of $NaPF_6$ and other diglymes .[197,206,207]

Overall, there is much scope for greater neutron utilisation to investigate SIB electrolytes. Like Jensen *et al*.[197] did, sodium salts can be investigated in various solvents using nPDF, especially the novel promising glyme types. Ionic liquid and solid state electrolytes could also benefit from structural neutron investigations such as with ND and nPDF as well as INS and QENS. The latter two spectroscopic techniques could be of great help to all electrolyte materials given the importance of sodium ion diffusion, which is directly related to conduction of electric charges, especially if they are accompanied with other techniques capable of dynamics investigation, such as NMR, EIS, and computational techniques. Furthermore, SEI has to be studied much further as it has a huge impact on battery performance,[16,196] and yet only few neutron studies have been done.[76,175] Other neutron techniques, such as SANS, are highly useful for this purpose, especially if done *in situ*. The greatest opportunity in this regard might be SEI in SSEs because there has been little research performed other than using neutrons,[16] and the lack of hydrogen in many of these materials means that they can be solved using complementary computational techniques, which is explained below.

## 3. Computational methods and neutron spectroscopy

Computational techniques are another brilliant set of tools for understanding the nature of SIB materials, being even more popular than neutron techniques. Unlike neutrons, which are available only in a few large-scale facilities worldwide, smaller simulations can be done on computers and the facilities required for large or long simulations are more accessible. They can also reveal the same and even deeper information than neutrons can, whether that be on structural, dynamic, or quantum properties. However, their accuracy is limited by the validity and precision of models used, and results should be validated by experiments.[226]

The most common techniques used in SIB research are density functional theory (DFT), molecular dynamics (MD), and Monte-Carlo (MC) simulations. DFT is a highly versatile *ab initio* (from first principle) computational method[227,228] were used to study phase[43,58,59,70,96,108,113,114,139] and other changes[105,141,155,165,168,169,171,178,188] observed during charge-discharge cycling, structure,[44,46,147,148,199,211,216,229,58,63,70,103,106,114,139,141] diffusion,[42,43,168,169,208,216,222,56,93,114,134,140,143,147,155] and to calculate electronic and vibrational density of states (DOS)[42,44,202,51,59,71,104,116,165,199,200] as well as other properties.[56,75,76,106,114,136] MC modelling used repeated (semi-)random sampling to evaluate structures[73,134] and energies. It also has been used to expand DFT results, such as to a larger system,[108] or higher voltage.[143] MD simulations apply Newtonian dynamics to all atoms, modelling their movement, and are mostly used to examine diffusions. They can either use a classical potential[173,193,213,223] or a quantum-mechanical one,[46,73,208,216,218,221,223,225,75,81,96,139,175,184,205,207] in which case they are referred to as *ab initio* MD (AIMD).

A comprehensive review of computational techniques used in SIB materials is beyond the scope of this article, for that the reader is pointed towards earlier reviews;[230,231] instead we are interested in how they are used alongside neutron techniques. Among one third of all work, where neutron have been used to investigate SIB materials, used DFT, MD, or MC simulations, with DFT being the most widely used and MC the least. However, most of these studies used the computational and neutron techniques independently of each other.

This is especially the case if the neutron technique used is ND. Since ND is usually employed only as a complement to XRD for structural information, there are few notable cases where it supports and validates a simulation. The simplest ones are when computationally predicted structures have been compared with ND and its refinements,[63,73,103,139,141,143,148,211,216,229] and vice-versa.[116,218,219] Similarly, simulated phase changes can be compared with *in situ* ND.[59] Rather than comparing geometries, it is also possible to compute a scattering function and compare it to one obtained from ND.[221] Lastly, simulated magnetic structure is compared with magnetic structure obtained from ND experiments.[106,114,136]

Calculations of PDF from MD simulations are more common [175,193,221] than calculations of nPDF.[207] Apart from bringing out the unique insight provided by a PDF, MD simulations can be used to validate interatomic force fields used for these calculations. The MD simulations also have been used to investigate INS spectra and the dynamics of water in MXene type anode materials.[193]

INS or neutron vibrational spectroscopy (NVS) is the only neutron technique that is almost always accompanied by a complementary computational simulations. Lattice dynamics simulations using DFT can generate vibrational DOS, which can be verified with INS experiments. [184,199,200,202,206,207] Generalised vibrational densities of states (GVDOS) also can be extracted from an INS spectra and compared with vibrational densities states (VDOS) obtained from DFT simulations.

Last but not least, QENS and MD can be used together to get insight into dynamical processes in SIB materials. However, less than half of all publications attempt to do so,[73,75,184,205,223,225] , where below 10% have used DFT[208]. Both QENS and MD are capable of revealing information about long- and short-range diffusion, and so if they are used together, QENS results can provide verification of MD results, allowing MD to be used on a (de)intercalated system,[75] a system with more or fewer defects or vacancies,[73,184,225] or to obtain other information inaccessible to QENS.[75,225] Conversely, QENS can be used to complement the short simulation length of AIMD,[205] and it can also provide results at very low temperatures.[75,205,225] Having said that, information can also be carefully extracted from AIMD to supplement a QENS result with having often a large background.[184,223,225]

Beyond these, another simulation method, the maximum entropy method (MEM) - has been shown to be highly effective as a computational technique that can calculate various structural information as well as visualise them from diffraction data, primarily ND in this current context. It can be used simply to extend ND to further understand a material's structure,[81,88,131,216] but it can also reveal a lot of information about sodium ion diffusion,[164,213,216,219] which is especially powerful when paired with DFT or MD.[213,216,219] Wiedemann *et al.*[164] even managed to extract migration barriers from it, though they did not use a simulation for comparison. Overall, MEM is a powerful technique that allows for maximum utilisation of diffraction data, so it is surprising that it is not used more widely, particularly to correlate QENS and INS results.

Computations based on the bond valence (BV) model[232] similarly use refined diffraction data to calculate and visualise various structural information. ND[80,130,148,189,190,213,216] and joint ND and XRD refinements[134,137,150] , though the source of data is often not explicitly mentioned.[42,59,132,136,143,151,188,222] BV site energy calculation[233] can determine diffusion activation energy.[188] Sometimes, BV sums are simply listed,[134,136,150] but their values are not transferrable to study structure[130,151] or to determine oxidation states[137,148,151,189] or any other properties.[222] Further, by sampling the unit cell, a BV sum map[232] can be created to study diffusion and visualise migration pathways.[132,137,143] A variation, BV energy landscape, can compute similar results,[42,59,80,190,213,216] but it has also been used for structural analysis.[216] As such, using either of these latter techniques alongside MEM[213,216] or computational techniques[143,213,216] can yield valuable results. There is no report in our knowledge, where BV method has been used to analyse QENS results in SIB research.

# 4. Conclusions and Future Outlook

Despite the booming research interest in SIB materials, the use of neutron techniques to investigate these materials are just emerging. The most used neutron techniques regarding this is ND. In most cases, however, ND is performed to solve a particular challenge alongside with XRD. Further, despite the range of materials it has been applied to, there are whole classes, such as cathode material PBAs, hardly studied with it. Although many layered oxides as cathode materials are well studied by ND, there are gaps which would get benefit from using ND. Arguably the worst of all, however, is that the maximum amount of information is rarely extracted from ND resulting this expensive technique has not been used up to its potential.

Other broadly used neutron techniques is neutron PDF. This has been used in various materials, particularly in carbonaceous anodes and amorphous electrolytes, but only to a small degree in cathode materials. It can be potentially useful to apply this technique across all materials suitable for SIBs, particularly alongside with MD simulations.

The use of neutron spectroscopies, mainly INS and QENS are still rare, although the insight of dynamics is essential for designing SIB materials. QENS experiments have been done mostly on electrolytes, even though it has a wide range of applicability on all types of materials. This is mainly due to unavailability of very high resolution neutron spectroscopic instruments are required to probe the slow dynamics in electrode materials. This technique has potential to provide a wealth of information about microscopic diffusion. When QENS is used in conjecture with ND, a structure dynamics correlations can be revealed. Adopting it more widely could bring a fundamental understanding of diffusion in SIB materials closer, especially if it is accompanied by simulations using MD, MEM and BV methods.

Simulations using both first principle- based DFT and force field based methods are widely used across all SIB materials. However, use of these powerful techniques to explain neutron spectroscopic or diffraction data is still rare. There is a huge scope to fill this gap by combining these two techniques.

Experiments in operando condition, such as ex-situ and in-situ methods with neutrons, are still lagging behind. Phase changes during charging-discharging cycling is of utmost importance in many materials, especially layered oxides. For this purpose, improvement of suitable sample environment to be used in *ex situ*, *in situ*, and *operando* methods for neutron spectroscopy is the utmost important. Lastly, use of SANS is understandably rare, but once greater emphasis is placed on SEIs. This technique has great potential to be utilised more.

Overall, while neutron techniques are increasingly used in SIB research, they are not using their full potential. Cathode materials are the most well studied with neutrons, but even layered oxides and polyanionic compounds can benefit from wider adoption, let alone PBAs and organic cathodes. Many types of anodic materials have not been studied with neutrons at all, and those that have, have been only little. Solid electrolytes have seen some attention, though it should continue, but other types need neutrons. We hope that improvements instrumentation, sample environments and simulations methodology allow these techniques to be more accessible in future.